\documentclass[10pt]{bmc_article}    

\usepackage{cite} 
\usepackage{url}  
\usepackage{ifthen}  
\usepackage{multicol}   
\usepackage[utf8]{inputenc} 
\usepackage{graphicx}
\usepackage{amsmath}

\newboolean{publ}
\def\R{\mathcal{R}}
\def\Q{\mathcal{Q}}
\def\H{\mathcal{H}}
\newtheorem{theorem}{Theorem}
\newtheorem{lemma}{Lemma}
\newtheorem{corollary}{Corollary}

\newenvironment{bmcformat}{\begin{raggedright}\baselineskip20pt\sloppy\setboolean{publ}{false}}{\end{raggedright}\baselineskip20pt\sloppy}

\begin{document}
\begin{bmcformat}

\title {Autocatalytic Sets Extended: Dynamics, Inhibition, and a Generalization}
\author{Wim Hordijk\correspondingauthor$^1$%
         \email{Wim Hordijk\correspondingauthor - wim@WorldWideWanderings.net}
       and 
         Mike Steel$^2$%
         \email{Mike Steel - mike.steel@canterbury.ac.nz}%
       }
\address{%
    \iid(1)SmartAnalytiX.com\\
    \iid(2)Biomathematics Research Centre, University of Canterbury, Christchurch, New Zealand
}%
\maketitle

\begin{abstract}
{\bf Background:} Autocatalytic sets are often considered a necessary (but not sufficient) condition for the origin and early evolution of life. Although the idea of autocatalytic sets was already conceived of many years ago, only recently have they gained more interest, following advances in creating them experimentally in the laboratory. In our own work, we have studied autocatalytic sets extensively from a computational and theoretical point of view.\\
{\bf Results:} We present results from an initial study of the dynamics of self-sustaining autocatalytic sets (RAFs). In particular, simulations of molecular flow on autocatalytic sets are performed, to illustrate the kinds of dynamics that can occur. Next, we present an extension of our (previously introduced) algorithm for finding autocatalytic sets in general reaction networks, which can also handle inhibition. We show that in this case detecting autocatalytic sets is fixed parameter tractable. Finally, we formulate a generalized version of the algorithm that can also be applied outside the context of chemistry and origin of life, which we illustrate with a toy example from economics.\\
{\bf Conclusions:} Having shown theoretically (in previous work) that autocatalytic sets are highly likely to exist, we conclude here that also in terms of dynamics such sets are viable and outcompete non-autocatalytic sets. Furthermore, our dynamical results confirm arguments made earlier about how autocatalytic subsets can enable their own growth or give rise to other such subsets coming into existence. Finally, our algorithmic extension and generalization show that more realistic scenarios (e.g., including inhibition) can also be dealt with within our framework, and that it can even be applied to areas outside of chemistry, such as economics.
\end{abstract}

\ifthenelse{\boolean{publ}}{\begin{multicols}{2}}{}

\section*{Background}

The idea of {\it collectively autocatalytic sets} has been introduced more or less independently several times \cite{Kauffman:71,Eigen:77,Dyson:82}, and was subsequently used in a number of origin of life models \cite{Wachterhauser:90,Ganti:97,Rosen:91,Letelier:06}. Recent experimental advances in creating such sets in the laboratory \cite{Sievers:94,Ashkenasy:04,Hayden:08,Taran:10} have generated a renewed interest in autocatalytic sets. And there is growing evidence that simple autocatalytic cycles may indeed have been at the core of the origin of life \cite{Braakman:12}.

In our own work, we have studied autocatalytic sets extensively from a computational and theoretical point of view  \cite{Steel:00,Hordijk:04,Mossel:05,Hordijk:10,Hordijk:11,Hordijk:12,Hordijk:12a}. We briefly review some of the main definitions and results here. First, we define a {\it chemical reaction system} (CRS) as a tuple $\Q=\{X,\R,C\}$ consisting of a set of molecule types $X$, a set of reactions $\R$ (transforming reactants to products), and a catalysis set $C$ indicating which molecule types catalyze which reactions. We also include the notion of a food set $F \subset X$ of molecule types assumed to be freely available from the environment. In a particular model of a CRS, known as the binary polymer model \cite{Kauffman:71,Kauffman:86,Kauffman:93}, molecule types are represented as bit strings up to a certain length $n$, reactions are simply ligation and cleavage, and catalysis is assigned at random according to some parameter $p$ (the probability that a given molecule type catalyzes a given reaction). The food set consists of all molecule types up to a certain length $t \ll n$.

Informally, an {\it autocatalytic set} that is self-sustaining (or an RAF set, in our terminology) is now defined as a subset $\R' \subseteq \R$ of reactions (and associated molecule types) in which:
\begin{enumerate}
\item each reaction $r \in \R'$ is catalyzed by at least one molecule type involved in $\R'$;
\item all reactants in $\R'$ can be produced from the food set $F$ by using a series of reactions only from $\R'$ itself.
\end{enumerate}
A formal definition is provided in \cite{Hordijk:04,Hordijk:11}, where we also introduced a polynomial-time (in the size of the reaction set $\R$) algorithm for finding RAF sets in a general CRS. Note that our framework is somewhat different from that of \cite{Andersen:12}, for which it was shown that maximizing the output flow and recognizing autocatalysis is NP-complete.

Some of our main results are that autocatalytic sets are highly likely to exist, even at very moderate levels of catalysis. For example, in the binary polymer model, each molecule needs only catalyze between one and two reactions, on average, to have a high probability of RAF sets emerging \cite{Hordijk:04,Mossel:05}. Also, more realistic assumptions, such as template-based catalysis (as opposed to merely random catalysis) can be built into the framework easily, and do not significantly change the results \cite{Hordijk:11}. In fact, required levels of catalysis for RAF sets to form in the template-based model can be predicted analytically from the (known) required levels in the base (random) model \cite{Hordijk:12}. And finally, RAF sets can often be decomposed into smaller RAF subsets (possibly even exponentially many), which can provide a mechanism for the evolvability of autocatalytic sets \cite{Vasas:12,Hordijk:12a}.

Here, we continue our studies of autocatalytic sets with various extensions of our framework. First, we investigate actual dynamics of autocatalytic sets. We present some initial but insightful results from simulating molecular flow on RAF sets. Next, we present an extension of our algorithm for detecting autocatalytic sets when inhibition is also considered, i.e., molecules that can potentially prevent a reaction from happening. In an earlier paper we proved that the general problem of detecting autocatalytic sets when inhibition is present, is NP-complete \cite{Mossel:05}. However, here we show that the problem is actually fixed parameter tractable, i.e., if the number of inhibiting molecules is not too large, autocatalytic sets (or their absence) can still be determined in polynomial time. Finally, in a recent paper we speculated about a generalized theory of autocatalytic sets beyond the context of chemistry and origin of life \cite{Hordijk:12a}. Here, we make a first concrete step in this direction by formulating a generalized version of our RAF algorithm which does not depend on the specifics of chemistry (i.e., molecules and reactions), and can be applied in a more general setting. These results are presented, in three parts, in the following section.

\section*{Results and discussion}

\subsection*{Part I: Dynamics}

In our work so far, we have mostly looked at autocatalytic sets in terms of their graph theoretical properties. However, this has ignored dynamics, i.e., actual molecular flow on autocatalytic sets. Here, we fill this gap by presenting initial results on studying the dynamics of RAF sets. In particular, we provide two examples, a constructed one and a realistic one, to show several aspects of the molecular flow that (can) occur. To a large degree, these dynamical results confirm what had already been analyzed, concluded, and speculated in our earlier (structural) studies, but they also shed some new light on autocatalytic sets and their behavior. Note that a related dynamical study was reported recently \cite{Filisetti:11}, although here we focus more directly on the actual molecular flow on RAF sets themselves.

\subsubsection*{A constructed example}

Consider the simple chemical reaction system (CRS) $\Q=\{X,\R,C\}$ within the binary polymer model, of which the reaction graph is shown in Figure 1, and with a food set $F=\{00,01,10,11\}$. This CRS consists of four reactions, each one being a bi-directional ligation/cleavage reaction, either combining two food molecules into a unique molecule of length four (in the ``forward'', or ligation reaction), or splitting up a molecule of length four into two food molecules (in the ``backward'', or cleavage reaction). The two reactions at the top are mutually catalyzed by each others ligation product, and form a 2-reaction autocatalytic (RAF) set. The two bottom reactions are not catalyzed, and are thus not part of any RAF set. However, these two sets of reactions (the top RAF one and the bottom non-RAF one) compete with each other for the food molecules.

Using the Gillespie algorithm \cite{Gillespie:76,Gillespie:77}, we simulate the flow of molecules on this constructed reaction graph. Food molecules are assumed to be always available, and are kept at a minimum concentration of five molecules each (i.e., if after one of the ligation reactions the concentration of a food molecule has dropped below five, it is immediately replenished). One rationale for this is that the reaction system can be assumed to be ``contained'' inside some compartment, for example a lipid layer \cite{Segre:01} or simply naturally occurring cavities in the soil \cite{Martin:03,Martin:07}. So, even though the food molecules are in ``unlimited'' supply in the environment, they still need to be taken up and brought inside the compartment to be used as reactants.

The presence of a catalyst increases the probability that a reaction will happen in direct proportion to the catalyst's current concentration. However, with this constructed example we are specifically interested in the effects of {\it auto}catalysis, and we ignore the fact that a catalyst normally also increases the basic reaction rate. So, for this example, the reaction rates of catalyzed and uncatalyzed reactions are kept equal (we relax this assumption again in the more realistic example in the next subsection).

To confirm that the simulation produces correct results, we first consider the reactions as uni-directional ligation reactions only. In this case, we expect a linear growth rate over time in the concentrations of the products $0011$ and $0110$ of the bottom two (non-RAF) reactions, but an exponential growth rate in the concentrations of the products of the top two (RAF) reactions, given that they form an autocatalytic set. Figure 2 shows the results, and indeed confirms this expectation (note that the y-axis is on a log-scale, so the exponential growth shows as a straight line). Since this is a simple model setting, the time units (x-axis) are arbitrary.

Next, we consider the full system, including the ``backward'' (cleavage) reactions. In this case, the molecule concentrations cannot grow unlimited, as they start breaking down at a rate proportional to their concentration. So, one would expect them to reach some equilibrium distribution. Figure 3 shows the result (simulating 10,000 reaction events). As expected, the molecular concentrations do indeed seem to reach an equilibrium distribution (instead of unlimited growth as with the uni-directional reactions in Figure 2). However, the two reactions forming an RAF set still have a large advantage over the two non-RAF reactions. The growth rate in concentrations of the molecules $0001$ and $1011$ (red and green lines) is much higher (until it levels off) than that of the molecules $0011$ and $0110$ (blue and purple lines). Also, the RAF set is able to maintain a much higher concentration of its ligation products than the non-RAF set. The light blue line shows the concentration of one of the food molecules over time (for reference). The concentrations of the other food molecules are similar due to the symmetry in the system.

This result clearly shows that the advantage of RAF sets over non-RAF sets is due to the particular, catalytically closed, structure of an RAF set. Even if uncatalyzed reactions have the same (basic) reaction rate as catalyzed reactions, as in this simulation, RAF sets still outcompete non-RAF sets due to the self-reinforcing autocatalytic feedback. However, the equilibrium distribution that is reached does depend largely on the ratio of the reaction rates between the ligation and the cleavage reactions. If this ratio is large enough, the concentrations of the product molecules can be maintained at a high level, as in Figure 3. However, reducing this ratio causes the level of the equilibrium concentrations to drop, until at some point there is no advantage anymore for the RAF set over the non-RAF set. Figure 4 shows such a situation (again simulating 10,000 reaction events).

\subsubsection*{A realistic example}

Next, we consider an example of an actual autocatalytic (RAF) set that was found by our RAF algorithm in an instance of the binary polymer model with $n=5$, $t=2$, and $p=0.0045$. Figure 5 shows this RAF set, which consists of eight bi-directional (ligation/cleavage) reactions. The food set is $F=\{0,1,00,01,10,11\}$.

This maximal RAF set actually consists of several RAF subsets (in \cite{Hordijk:12a} we show formally how RAF sets can be decomposed into, possibly exponentially many, RAF subsets). First there are two simple (1-reaction) irreducible RAF sets contained inside the yellow and purple boxes, respectively. Given that their reactants and catalysts are all food molecules, these subRAF will always be present. Then there is the 3-reaction subRAF contained inside the red box. This subRAF actually includes the purple (1-reaction) irrRAF, but can only ``grow'' into the full 3-reaction red subRAF once molecule type $1010$ is present. This molecule type catalyzes its own ligation from two instances of the food molecule $10$, so this reaction will have to happen spontaneously (uncatalyzed) first, before the red subRAF can come into full self-sustaining existence (in fact, this reaction is actually an irrRAF in itself, but for the purposes of the dynamical analysis here, we do not consider it separately as such, as it immediately gives rise to the full red subRAF as soon as it comes into existence). Next, there is the 3-reaction irreducible RAF set contained inside the blue box. This blue subRAF also needs to be seeded, by one of the three reactions happening uncatalyzed (or one of the required molecules coming from elsewhere). Finally, there is the reaction contained in the green box, which strictly speaking is not an RAF by itself, but once molecule type $111$ (produced by the blue subRAF) is available, it can become an ``extension'' of the blue subRAF. However, since the green reaction is catalyzed by its own product, it also needs to happen spontaneously at least once, before it can maintain its own existence autocatalytically.

Using the Gillespie algorithm again, we now study the molecular flow on this maximal RAF of Figure 5. In this simulation, we do make a difference between the reaction rates of catalyzed and uncatalyzed reactions, to show the effect of some of the subRAFs needing to be seeded by spontaneous reactions. In particular, if for a given reaction the reactants are present but not the catalyst, the reaction can still go ahead uncatalyzed, but at a reduced rate. For the sake of the simulation, we used a small reduction factor of 20. A higher, more realistic, factor is of course possible, but does not change the qualitative results, and simply means we need somewhat larger time-scales to observe similar behavior. Figure 6 shows the concentrations over time (simulating 25,000 reaction events this time) for the products of the eight reactions making up the maximal RAF set.

The dynamics of the molecular concentrations are a direct reflection of the particular structure of the maximal RAF set in terms of its subRAFs. First of all, the concentrations of the products of the two 1-reaction irrRAFs (indicated, as in Figure 5, with yellow and purple lines, respectively), immediately start growing at a steady rate (although not exponentially, as they are catalyzed by food molecules, which remain in relatively low concentrations). However, the other subRAFs all need to be seeded by a spontaneous reaction. The first such event happens around time 0.3, when one of the reactions in the blue subRAF happens uncatalyzed. But once this has happened, the blue subRAF as a whole can come into existence and grow in concentration. Note that the two product types $010$ (solid blue line) and $11100$ (dashed blue line) immediately grow rapidly in concentration, but $111$ (dotted blue line) has a damped growth, as it is also used again as a reactant.

The next spontaneous event happens around time 0.5. Recall that around time 0.3 the molecule type $111$ came into existence, but for the green reaction to become an extension of the blue subRAF, it will still need to happen uncatalyzed at least once (given that it is catalyzed by its own product). However, when this happens (around time 0.5), the concentration of its product type $01111$ (green line), supported by a product of the blue subRAF, immediately starts to grow rapidly. Finally, a last required spontaneous event happens around time 0.55, when molecule type $1010$ is created, which then gives rise to the red subRAF coming into full existence (given that the purple irrRAF it contains was already present).

Some additional observations can be made about these dynamics. First, molecule type $00100$ (dashed red line), a product of the red subRAF, was actually already present before the full red subRAF came into existence, as a result of spontaneous (uncatalyzed) reactions. However, its concentration only really starts growing once molecule type $1010$ (its catalyst; solid red line) is present. Next, the concentration of the product of the purple irrRAF ($100$, purple line) starts decreasing again as soon as the red subRAF comes into existence, as this molecule type is used as a reactant within the red subRAF. And finally, note that the three molecule types that seem to grow in concentration without limit ($00100$, $11100$, and $01111$) are the ones that actually have a non-food molecule as one of their building blocks (reactants). Food molecules remain present in relatively low concentrations (although they are replenished when they fall below a concentration of five), but non-food molecules reach higher concentrations, and thus increase, in direct proportion to their concentration, the rate at which reactions that use them as reactants will happen. However, at some point the growth of these three molecule types also levels off, because of the backward (cleavage) reactions happening more and more often as well (similar to what happens in Figure 3); for readability of the graph, though, concentrations above 100 molecules are not shown in Figure 6.

These initial results are, of course, only a first step towards a more complete study of the dynamics of autocatalytic sets. However, they already provide some very useful and interesting insights into the kinds of dynamics one can observe in RAF sets, and also confirm some of the claims made recently on how subRAFs can enable their own growth and each others coming into existence \cite{Hordijk:12a}. Moreover, there are many directions in which such a dynamical analysis can be extended. For example, one can consider having autocatalytic (sub)sets enclosed in different compartments, able to grow and reproduce (once a threshold concentration of certain molecule types is reached). Variation can then be introduced by only passing on a (perhaps random) subset of the molecules from the parent to the offspring, i.e., offspring compartments can possibly have different combinations of existent subRAFs, enabling an evolutionary process to happen \cite{Vasas:12}. As another example, one can ask what will happen if there are {\it inhibitors} present in the system, i.e., molecules that can actually {\it prevent} a reaction from happening. In the next section, we describe an extension of our RAF algorithm for dealing with such a situation.

\subsection*{Part II: Inhibition}

Given a chemical reaction system, $\Q= (X, \R, C)$, with food set $F$, suppose we have a collection $(X_1, \R_1), (X_2, \R_2), \ldots, (X_k, \R_k)$
where $X_i \subset X$, and $\R_i \subset \R$. The interpretation of the pair  $(X_i, \R_i)$ is that every molecule $x \in X_i$ {\it inhibits} every reaction $r \in \R_i$. Notice that any pattern of inhibition can be represented this way, for example by numbering the reactions, and taking $\R_i = \{r_i\}$ and $X_i$ to be the set of molecules that inhibit $r_i$ (or we may number the molecules, and take $X_i = \{x_i\}$ and $\R_i$ to be the set of reactions inhibited by $x_i$). We wish, however,  to consider `types' of molecules that will inhibit `types' of reactions so that $k$ can be chosen to be not too large.

We say that a subset $\R' \subseteq \R$ forms an {\em uninhibited RAF}, or more briefly a $u-$RAF, if $\R'$ is an RAF (in the usual sense) and $\R'$ contains no reaction that is inhibited by any molecule that is involved in $\R'$. For a more formal definition, let ${\rm supp}(\R')$ denote the {\em support} of $\R'$ -- this is the set of molecules that are either reactants or products of reactions in $\R'$ (this is the same as the union of the set of molecules in $F$ that are reactants of reactions in $\R'$, and the set of products of reactions in $\R'$). Uninhibited RAFs are now defined more formally as follows.

{\bf Definition}: Given a chemical reaction system, $\Q= (X, \R, C)$, with food set $F$, a subset $\R'$ of $\R$ is  a $u-$RAF if
\begin{itemize}
\item[(u-1)] $\R'$ is an RAF.
\item[(u-2)] $\R' \cap \R_i \neq \emptyset \Rightarrow {\rm supp}(\R') \cap X_i = \emptyset.$
\end{itemize}
Note that any subset of a set of reactions that satisfies condition (u-2) also satisfies (u-2); this implies that any subset of a $u-$RAF that is an RAF is also a $u-$RAF.

Determining whether a CRS contains a $u-$RAF was shown to be an NP-complete problem in \cite{Mossel:05}. However, here we show that the problem is {\it fixed parameter tractable} in the parameter $k$. So, provided $k$ is not too large, we can still find $u-$RAFs in a CRS efficiently (or determine that a $u-$RAF does not exist).

We first require some additional definitions. Let $[k] := \{1,\ldots, k\}$, and for any subset $J$ of $[k]$, let 
\begin{equation}
\label{RJ1eq} 
\R^J := \{r \in \R: {\rm supp}(r) \cap X_j = \emptyset \mbox{ for all } j \not\in J\},
\end{equation}
and let
\begin{equation}
\label{RJ2eq} 
\R_J: = \{r \in \R, r \not\in \R_j \mbox{ for all } j \in J\}.
\end{equation}

In the following theorem, the set $\R^J\cap \R_J$ plays a prominent role (where $J$ is a subset of $[k]$); this is precisely the set of reactions $r$ in $\R$ for which (i) $r$ does not belong
to $\R_j$ for any $j \in J$ and (ii) if $r \in \R_{j'}$ (for some $j'  \not\in J$) then none of the molecules in the support of $r$ lie in $X_{j'}$.
Recall from \cite{Hordijk:12a} that for any subset $\R^*$ of reactions in $\R$, $s(\R^*)$ is the maximal subRAF contained within $\R^*$ (as computed by our RAF algorithm) or the empty set if no such subRAF of $\R^*$ exists. We can now state our first theorem.

\begin{theorem}
Given a chemical reaction system, $\Q= (X, \R, C)$, with food set $F$, the following assertions hold: 
\mbox{ }
\begin{itemize}
\item[(i)] For any subset $J$ of $[k]$, if $s(\R^J\cap \R_J)$ is non-empty, then it is a $u-$RAF.
\item[(ii)] If $\R'\subseteq \R$ is a $u-$RAF, then $\R' \subseteq \R^J \cap \R_J$ where 
\begin{equation}
\label{RReq}
J = \{j \in [k]:  {\rm supp}(\R') \cap X_j  \neq \emptyset\}.
\end{equation}
\item[(iii)] The set of maximal $u-$RAFs is precisely the collection of all non-empty subsets of $\R$ of the form $s(\R^J \cap \R_J)$ as $J$ ranges over subsets of $[k]$.
\end{itemize}
\end{theorem}

{\em Proof: }
For part (i) we know that if $s(\R^J\cap \R_J)$ is non-empty, then it is an RAF (from \cite{Hordijk:04}), thus it suffices to verify property (u-2) in the definition of a $u-$RAF above for the set
$\R^* := \R^J\cap \R_J$, which implies that $s(\R^*)$ will also satisfy property (u-2), since it is a subset of $\R^*$.

Suppose, to the contrary, that property (u-2) in the definition of a $u-$RAF is violated by $\R^*$, then we can derive a contradiction as follows. For some $i \in [k]$ we must have:
\begin{equation}
\label{Req}
\R^* \cap \R_i \neq \emptyset \mbox{  and  }  {\rm supp}(\R^*) \cap X_i \neq \emptyset.
\end{equation}
In particular, there exists a reaction, say $r_1$, in $\R^* \cap \R_i$. Moreover, since ${\rm supp}(\R^*) = \cup_{r \in \R^*} {\rm supp}(r)$, the second part of Eqn. (\ref{Req}) implies that there also exists a reaction, say $r_2$, in $\R^*$ for which ${\rm supp}(r_2) \cap X_i \neq \emptyset.$ Now, since $r_1 \in \R_J$ and $r_1 \in \R_i$ it follows, by the definition of $R_J$, that $i$ cannot be in $J$. Now consider $r_2$. This reaction is in $R^J$ and so, since $i$ does not lie in $J$, we must have ${\rm supp}(r) \cap X_i = \emptyset$. But this contradicts the choice of $r_2$. This establishes part (i).

For part (ii), suppose that $\R' \subseteq \R$ is a $u-$RAF. It suffices to show that $\R' \subset \R^J$ and that $\R' \subseteq \R_J$ for the set $J$ described in Eqn. (\ref{RReq}); it follows that $\R'$ will be contained in the intersection of these two sets.

Observe that, for the set $J$ as described in Eqn. (\ref{RReq}), $\R_J$ is the set of reactions in $\R$ which do not lie in $\R_j$ for any $j$ for which ${\rm supp}(\R') \cap X_j \neq \emptyset$. Now, if $\R'$ is a $u-$RAF then by condition (u-2) in its definition, any reaction $r \in \R'$ must belong to $\R_J$. Similarly, for the choice of $J$ as described, $\R^J$ is the set of reactions $r$ in $\R'$ for which ${\rm supp}(r) \cap X_i$ is empty for all $i$ for which ${\rm supp}(\R') \cap X_i =\emptyset$, and so any reaction $r \in \R'$ must also lie in $\R^J$. This establishes the required two containments, and so part (ii).

For part (iii), we have shown by part (i) that non-empty sets of the form $s(\R^J\cap \R_J)$ are $u-$RAFs, so we need to check that all maximal $u-$RAFs are of this form. Suppose that $\R'$  is a maximal $u-$RAF. Then by part (ii) we know that $\R' \subseteq \R^J \cap \R_J$ for the choice of $J$ given by Eqn. (\ref{RReq}). Now, $s(\R') = \R'$ and so, by part (ii) $s(\R^J \cap \R_J)$ is a $u-$RAF containing $\R'$, and, since $\R'$ is assumed maximal, these two $u-$RAFs must coincide. Part (iii) now follows.

\begin{corollary}
Given a chemical reaction system, $\Q= (X, \R, C)$, with food set $F$, together with a family $\{(X_i, \R_i): i \in [k]\}$ of inhibition pairs, there is an algorithm for constructing one (or all) maximal $u-$RAFs (or determining that no $u-$RAF exists) in time $2^k p(n)$ where $p$ is a polynomial in the size $n$ of $\Q$.
\end{corollary}
{\em Proof:} Simply apply the RAF algorithm to compute $s(\R^J\cap \R_J)$ for all $2^k$ subsets $J$ of $[k]$.

\bigskip

{\bf Remark:}  In contrast to ordinary RAFs, $u-$RAFs need not be  closed under union, i.e., if $\R'$ and $\R''$ are two $u-$RAFs then $\R' \cup \R''$ may fail to be a $u-$RAF. Thus, in  general, a CRS may have  several  maximal $u-$RAFs, while there is always a unique maximal RAF.

So, this extension of our algorithm shows that, even though the general problem of finding RAF sets under inhibition is NP-complete, we can still deal with specific situations (such as when the number of inhibitors is limited) in a relatively efficient way. In the next section, we formulate another extension, or rather a generalization, of our RAF algorithm, which indicates that it can also be applied to problems outside of the context of chemistry and origin of life.

\subsection*{Part III: A generalization}

The original RAF algorithm is specifically formulated in the context of chemical reaction systems. However, it is also possible to state the algorithm in a more generalized form. This may be useful for (i) understanding its relationship to other algorithms, and (ii) extending it in further directions, both within the context of chemical reaction systems as well as for other applications (e.g., in economics, as already speculated in \cite{Hordijk:12a}).

Suppose we have arbitrary (finite or infinite) sets $Y, W$, where $W$ has a partial order ($\leq$, for example, take $W$ to be the set of subsets of some set partially ordered by set inclusion; as discussed later, this applies in the RAF setting), and functions
$$f: 2^Y \rightarrow W \mbox{ and } g: Y \rightarrow W$$
(here $2^Y$ refers to the set of all subsets of $Y$). Consider the function:
$\psi: 2^Y \rightarrow 2^Y$
which is determined by $f$ and $g$ according to the following rule:
$$\psi(A) = \{y \in A: g(y) \leq f(A)\},$$
for each subset $A$ of $Y$. Note that $\psi(A) \subseteq A$, for all $A \in 2^Y$.
 
\bigskip
 
{\bf Definition:} We say that a subset $A$ of $Y$ is {\em $gf-$compatible} if it is non-empty and satisfies the property that $g(y) \leq f(A)$ for all $y \in A$.
 
\bigskip 

For a subset $A$ of $Y$, and $k\geq 1$, define $\psi^{(k)}(A)$  to be the result of applying function $\psi$ iteratively $k$ times starting with $A$.  Thus,  $\psi^{(1)}(A)=\psi(A)$ and for $k\geq 1$, $\psi^{(k+1)}(A)=\psi(\psi^{(k)}(A))$. Notice that the sequence $(\psi^{(k)}(A), k\geq 1)$ is a nested, decreasing sequence of subsets of $Y$, and so we may define
$$\overline{\psi}(A) :=\lim_{k\rightarrow \infty}\psi^{(k)}(A) = \bigcap_{k \geq 1} \psi^{(k)}(A)$$
which is a (possibly empty) subset of $Y$. Moreover, if $Y$ is finite, then $\overline{\psi}(Y) = \psi^{(k)}(Y)$ for some $k \leq |Y|$.
 
To state the main result of this section, we recall two more standard definitions. A set $A \in 2^Y$ is a {\em fixed point} of $\psi$ if $\psi(A)=A$; and $f$ is {\em monotone} if it satisfies the property $A_1 \subseteq A_2 \Rightarrow f(A_1) \leq f(A_2)$. 

\begin{theorem}
\label{absthm}
Given sets $Y,W$ where $W$ is partially ordered, together with functions $f: 2^Y \rightarrow W \mbox{ and } g: Y \rightarrow W$, the following hold:
\begin{itemize}
\item[(i)] The $gf-$compatible subsets of $Y$ are precisely the non-empty subsets of $Y$ that are fixed points of $\psi$;
\item[(ii)] $\overline{\psi}(Y)$ is $gf-$compatible, provided it is non-empty; moreover, it contains all $gf-$compatible subsets of $Y$ provided that $f$ is monotone. In particular, when $f$ is monotone, there exists a $gf-$compatible subset of $Y$ if and only if $\overline{\psi}(Y)$ is nonempty.
\end{itemize}
\end{theorem}
\bigskip

{\em Proof: } If a subset $A$ of $Y$ is non-empty and $A=\psi(A)$ then $A =  \{y \in A: g(x) \leq f(A)\}$ and so $A$ is $gf-$compatible. Conversely if $A \neq \psi(A)$ then since $\psi(A) \subset A$, there exists $y \in A$ so that $g(y)$ is not dominated by $f(A)$ in the partial order. Thus $A$ is not $gf-$compatible. This establishes Part (i).

For Part (ii), let $B=\overline{\psi}(Y)$. Then $\psi(B) = \psi(\overline{\psi}(Y)) = \overline{\psi}(Y) = B$, so, $\overline{\psi}(Y)$ is a fixed point of $\psi$, and so, by part (i), is $gf-$compatible provided $B$ is non-empty. Also, if $f$ is monotone, and $A_1 \subseteq A_2$, then $\psi(A_1)$ equals
$$ \{y \in A_1: g(y) \leq f(A_1)\} \subseteq \{y \in A_1: g(y) \leq f(A_2)\} \subseteq \{y \in A_2: g(y) \leq f(A_2)\}$$
and this last set is  $\psi(A_2),$ so $\psi$ is monotone as a function from $2^Y$ to the set $2^Y$ partially ordered under set inclusion. Thus, if $B'$ is any $gf-$compatible set then, by part (i), $B'$ is a fixed point of $\psi$ and so, since $B' \subseteq Y$, we have $B'=\psi(B') \subseteq \psi(Y)$ and, by iteration of $\psi$, $B' \subseteq \overline{\psi}(Y)$, as claimed. The remaining claim in part (ii) now follows directly.

\bigskip
 
{\bf An algorithm:} Theorem~\ref{absthm} has the following immediate consequence when $Y$ is finite, and $f$ is monotone. In this case, consider the following `$gf-$algorithm'. Starting with $Y$, compute the sequence 
$\psi^{(k)}(Y)$ until it stabilizes. If this set is empty, then report that no $gf-$compatible subset of $Y$ exists, otherwise output the stable set $\overline{\psi}(Y)$, which is the unique maximal $gf-$compatible subset of $Y$. Provided that for each subset $A$ of $Y$, and element $y \in Y$, the values $f(A)$ and $g(y)$ can be calculated in polynomial time in $|Y|$, this algorithm runs in polynomial time in $|Y|$.
Notice that the algorithm begins with the set $Y$ and iteratively removes subsets of elements, until eventually arriving at a non-empty set $\overline{\psi}(Y)$ from which nothing further can be removed, or until all the elements of $Y$ are eliminated. 

\subsubsection*{Relationship to the original RAF algorithm}

First a simple observation: If a reaction $r$ is catalyzed by $k \geq 1$ molecules, then we can replace it (formally) by $k$ copies of this reaction, each of which is catalyzed by just one of the $k$-molecules. This way we get a set of reactions, each of which is catalyzed by exactly one molecule. We can thus think of this catalyst as an additional reactant and so the reaction proceeds precisely if all the `reactants' are present -- formally this is cleaner than saying ``all the reactants and at least one catalyst are present''. In fact, the implementation of our RAF algorithm is actually based on this idea. We call this `cleaner'  version the {\em expanded CRS}, and the catalyst chosen for any given reaction the {\em nominated catalyst}.  In this expanded CRS, given a reaction $r$, let $\rho(r)$ denote the set of reactants plus the nominated catalyst of this reaction. We now describe how Theorem 2 and the $gf-$algorithm applies. 

Given a CRS $(X, \R, C)$ and food set $F \subseteq X$, take $Y$ to be the set of all reactions in the expanded CRS, and take $M=X$, the set of all molecules, take $W=2^M$, partially ordered under set inclusion.  For our choice of the function $f$ we set
$f(A) = {\rm cl}_A(F)$, where ${\rm cl}_A(F)$ is the closure of the food set $F$ under a subset $A$ of reactions in the expanded CRS; this is  the set of all molecules in $X$ that can be constructed from $F$ 
by repeatedly applying just those reactions that lie in $A$ (and allowing any reaction in $A$ to proceed even if the nominated catalyst is not present).  Finally, we set $g(r) =\rho (r)$ (in the expanded model, so $\rho(r)$ includes the nominated catalyst). Then the $gf-$compatible subsets of $Y$ correspond exactly to the RAFs in the expanded CRS under the recent modified definition of RAF \cite{Hordijk:11}, and $\overline{\psi}(Y)$ is just what we call $s(Y)$ (the maxRAF for the expansion $Y$ of $\R$). Theorem~\ref{absthm}(ii) asserts this maxRAF can be found by the $gf-$algorithm, which is just the
 modified RAF algorithm \cite{Hordijk:11} applied in the expanded CRS, and the fact this RAF is the unique maximal RAF follows from the fact that the function $A \mapsto {\rm cl}_A(F)$ is monotone in $A$. 

The connection described assumes that we are working within the expanded CRS setting. However, we can easily relate this back to the original CRS setting by noting that if $A$ is a set of reactions, and $A'$ is the expanded version (replacing each reaction by $k$ copies each with a unique nominated catalyst)  then ${\rm cl}_A(F)$ (in the original setting) coincides with ${\rm cl}_{A'}(F)$ (in the expanded setting). Moreover, (i) for any RAF $A$ in the original setting, in the expansion of $A$ there is a subset (selecting an appropriate nominated catalyst for each reaction) that is an RAF in the expanded CRS, and (ii) for any RAF $A'$ in the expanded CRS, replacing the nominated catalyst of each reaction by its full complement of catalysts returns an RAF $A$ in the original CRS.

Notice that, apart from the monotonicity of the function $f(A)= {\rm cl}_A(F)$, a major factor that helps in guaranteeing a polynomial-time algorithm in the RAF setting is that $f(A)$ can be computed efficiently.

\subsubsection*{Novel and alternative applications}

We now present a simple application of Theorem~\ref{absthm} in a toy economic setting. Suppose $Y$ is a collection of individuals, each of whom produces or consumes different types of ``goods'', labeled $1,2,\ldots k$. For an individual $y \in Y$, let $g_i(y)$ be the maximum price individual $y$ is able to pay for good $i$ and let $f'_i(y)$ be the minimal price for which individual $y$ is willing to produce good $i$. To allow greater generality, if individual $y$ does not need good $i$ we can just set $g_i(y)=0$ and if individual $y$ does not produce good $i$ we can just set $f'_i(y) = \infty$. We assume that individuals can produce and sell as many goods as they wish (i.e. the individuals who are buying are not competing for a fixed number of items from any one seller).
 
We define a subset $A$ of $Y$ as {\em viable} if (i) it is non-empty, and (ii) every individual can afford to buy each good they need from at least one individual in $A$. We can formalize this as a $gf-$compatibility condition as follows.

Let $W= ({\bf R\cup\{{\infty}}\})^{k}$ (i.e., $k$-dimensional Euclidean space with infinity added to each co-ordinate) partially ordered in the usual way: $(x_1,\ldots, x_k) \leq (y_1, \ldots, y_k)$ if and only if $x_i \leq y_i$ for all $i$. Note that in this example $W$ is not a collection of subsets of a set (as in the RAF setting). Further, let $g(y) := (g_1(y), \ldots, g_k(y)) \in W$, and for a set $A$ individuals (i.e. $A \in 2^Y$) let
$$f(y) := (\max_{y \in A} f'_1(y), \cdots, \max_{y \in A}f'_k(y)) \in W.$$
Then a subset $A$ of $Y$ is viable precisely if for each $i$ and each $y \in A$, $g_i(y) \leq \max\{f'_j(A): j=1,\ldots, k\}$, which is equivalent to $g(y) \leq f(A)$ for all $y \in A$. In other words, $A$ is viable if and only if $A$ is $gf-$compatible. Moreover, notice that $f$ is monotone, and so Theorem~\ref{absthm}(ii) applies, so if there is a stable set, then there is a unique maximal one, and it can be found in polynomial time in the size of the population, by using the $gf-$algorithm.

This provides a (simple) example of how the $gf-$algorithm can be applied in other contexts, such as economics. This is a first concrete step towards a generalized theory of autocatalytic sets, as we recently proposed \cite{Hordijk:12a}.

As a further, and rather different, application we point out that the $gf-$algorithm also provides a polynomial-time solution to HORN-SAT, which is a basic problem in propositional logic, of deciding whether a given conjunction of Horn clauses is satisfiable \cite{pap:93}. Recall that a {\em Horn clause} is a clause with at most one positive literal, and any number of negative literals (a literal being a boolean variable which can be either `true' or `false'). HORN-SAT is of interest as it is `P-complete' (i.e. not only is it  in the complexity class P of problems having polynomial-time solutions, but {\em every} problem in the complexity class P can be reduced to HORN-SAT).

Suppose then, that we have an instance of HORN-SAT consisting of a conjunction of a set $\H$ of $n$ HORN clauses. Without loss of generality we will assume that  not all the clauses in $\H$ contain a positive literal, as this is equivalent to the condition that assigning each literal the truth value  `true' satisfies every clause in $\H$, and this can be easily checked. We indicate this restriction by saying that $\H$ is a {\em proper} instance of HORN-SAT.   Now we define the sets and functions we will use in the generalized RAF set-up.  We take $W=2^\H$  with the usual partial order on subsets.   Let $Y$ denote the set of all literals appearing in at least one clause in $\H$ (as a positive or negative literal).  For a subset $A$ of $Y$ let $f(A)$ be the set of clauses in $\H$ that contain at least one element of $A$ as a negative literal.  For $y \in Y$, let $g(y)$ be the set of clauses in $\H$ which either contain $y$ as a positive literal or else do not contain any positive literals. The following connection with $gf-$compatibility is established in the Appendix.

\begin{lemma}
\label{lemm}
For a proper instance $\H$ of HORN-SAT, a subset $A$ of $Y$ is $gf-$compatible if and only if the following truth assignment  satisfies every clause in $\H$:
\begin{equation}
\label{eqfalse}y ={\rm false } \Leftrightarrow y \in A
\end{equation}
\end{lemma}

By Lemma~\ref{lemm}, and the fact that $f$ is monotone, we can invoke Theorem 2(ii) and deduce that the $gf-$algorithm determines whether or not a proper instance of HORN-SAT has a satisfying assignment, and if it does, it will construct the  truth assignment that has a minimal set of literals set to `true'. This may all seem rather technical and irrelevant to chemistry, but it actually shows that an algorithm that was inspired by and constructed for solving a chemical problem in the context of the origin of life (finding autocatalytic sets in chemical reaction systems), turns out to be capable (in its generalized form) of solving {\it any } problem that is within the problem class P. So, perhaps this could lead to another application of {\it molecular computation} \cite{Adleman:94}.

\section*{Conclusions}

In our previous work, we already showed (both computationally and theoretically) that autocatalytic (RAF) sets are highly likely to exist. However, most of these results were based on graph theoretical properties of RAF sets. Here, we have shown that also in terms of dynamics such sets are indeed self-sustainable and can outcompete non-autocatalytic sets. Furthermore, these dynamical results confirm arguments made previously \cite{Hordijk:12a} about how RAF subsets can enable their own growth or give rise to other such subsets coming into existence.

Next, the extension described here of our RAF algorithm shows that more realistic scenarios (such as including inhibition) can also be dealt with within our framework. Despite the fact that the general problem of finding RAF sets when inhibition is present is NP-complete, in specific cases (such as when the number of inhibitors is not too large) it is still possible to detect RAF sets efficiently, due to our proof of this problem being fixed parameter tractable.

Finally, the generalization of our RAF algorithm shows that it can even be applied to areas outside of chemistry and origin of life, such as economics. This is an important first step towards a generalized theory of autocatalytic sets, as proposed in \cite{Hordijk:12a}. And, perhaps, it could lead to another application of molecular computation.

Of course there are still many further extensions possible. In terms of dynamics, a next step could be to consider multiple, possibly competing, compartments each having some (different) combination of subRAFs existent within them. This could then give rise to an evolutionary process along the lines of \cite{Vasas:12}. Also, it would be interesting to find further applications of the $gf-$algorithm outside of chemistry. We hope to work on some of these further extensions and generalizations in the future.

\section*{Competing interests}

The authors declare that they have no competing interests.

\section*{Authors' contributions}

WH implemented the algorithms (RAF and Gillespie) and performed the simulations and analysis. MS formulated and proved the mathematical theorems and algorithm extension and generalization. Both authors wrote the paper and approved the final version.

\section*{Acknowledgements}
\ifthenelse{\boolean{publ}}{\small}{}
MS thanks the Royal Society of New Zealand for funding support. We thank Stuart Kauffman for helpful and stimulating discussions.

{\ifthenelse{\boolean{publ}}{\footnotesize}{\small}
\bibliographystyle{bmc_article}
\bibliography{JSysChem}}

\ifthenelse{\boolean{publ}}{\end{multicols}}{}

\newpage
\section*{Appendix: Proof of Lemma~\ref{lemm}}
First suppose that $A$ is $gf-$compatible, and the truth assignment is as specified. Consider clause $c \in \H$. There are three possibilities:
\begin{enumerate}
\item If $c$ contains a positive literal that is not in $A$ then $c$ is satisfied, since that positive literal is assigned the value `true' under (\ref{eqfalse}).
\item If $c$ contains a positive literal $y$ in $A$ then $c \in  f(A)$ (since $g(y) \subseteq f(A)$, as $y \in A$), and so $c$ is satisfied under (\ref{eqfalse}).
\item If $c$ contains no positive literal, then $c$ is contained in $g(y)$ for {\em any} $y \in A$ (and there exists at least one such $y$ since $A$ is non-empty), and so the condition $g(y) \subseteq f(A)$ (for $y \in A$) implies, once again, that $c$ lies in $f(A)$, and so $c$ is satisfied under (\ref{eqfalse}).
\end{enumerate}
Thus all clauses in $\H$ are satisfied.  

Conversely, suppose the truth assignment determined by some set $A$ according to (\ref{eqfalse}) satisfies every clause in $\H$. Then $A$ cannot be the empty-set, otherwise every clause in $\H$ contains a positive literal, so $\H$ would not be proper. We wish to show that $g(y) \subseteq f(A)$ for all $y \in A$. Consider clause $c \in g(y)$. Then, by definition of $g$, either (i) $c$ has no positive literal, or (ii) $c$ has a positive literal and it is $y$, which lies in $A$. In case (i), the assumption that $c$ is satisfied implies that at least one of the negative literals in $c$ is set to false, which means one of these literals must be in the set $A$. Consequently $c \in f(A)$. Similarly, in case (ii), since the positive literal $y \in A$ is set to `false' at least one of the negated literals in $c$ must be set to false, which again requires this literal to lie in $A$, and hence $c \in f(A)$.
Thus $g(y) \subseteq f(A)$ for all $y \in A$, as required.

\bigskip

\newpage

\section*{Figures}

\subsection*{Figure 1 - A constructed example reaction graph}
The reaction graph of the constructed CRS with two sets of reactions: an RAF set (top two reactions) and a non-RAF set (bottom two reactions).\\
\includegraphics[width=6cm]{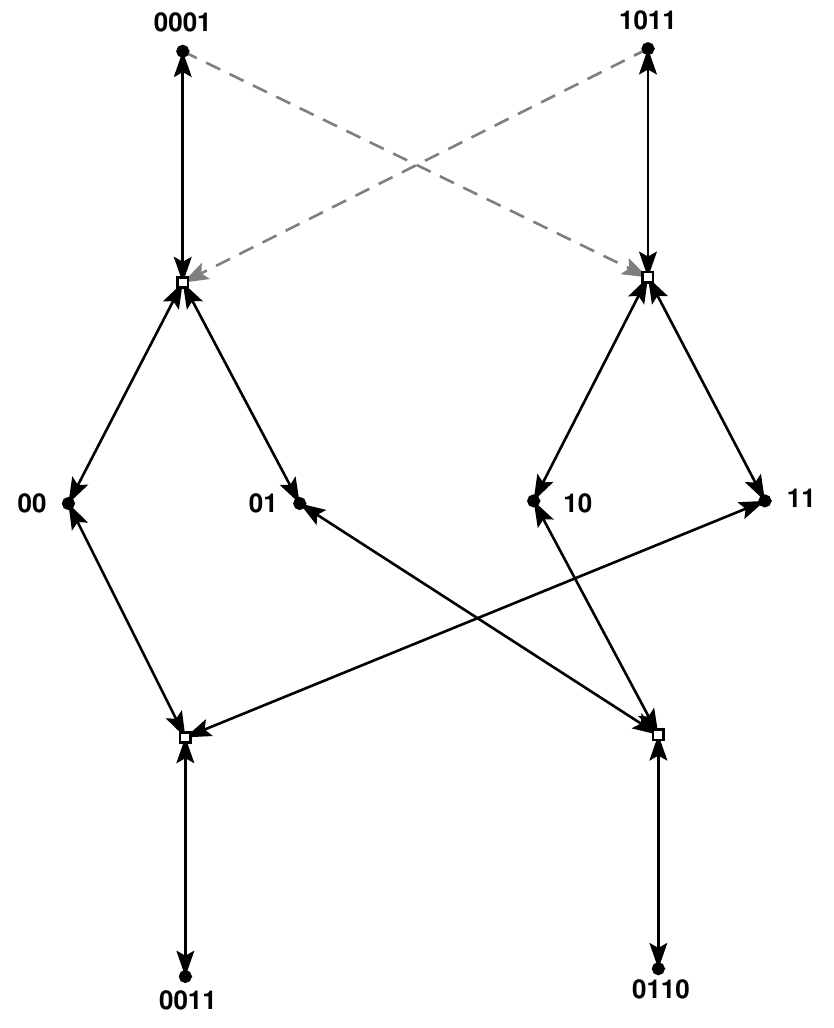}
\newpage
\subsection*{Figure 2 - Dynamics on the ligation-only reaction graph}
The molecular concentrations over time for the products of the four reactions in the constructed example CRS when only ligation reactions are considered.\\
\includegraphics[width=15cm]{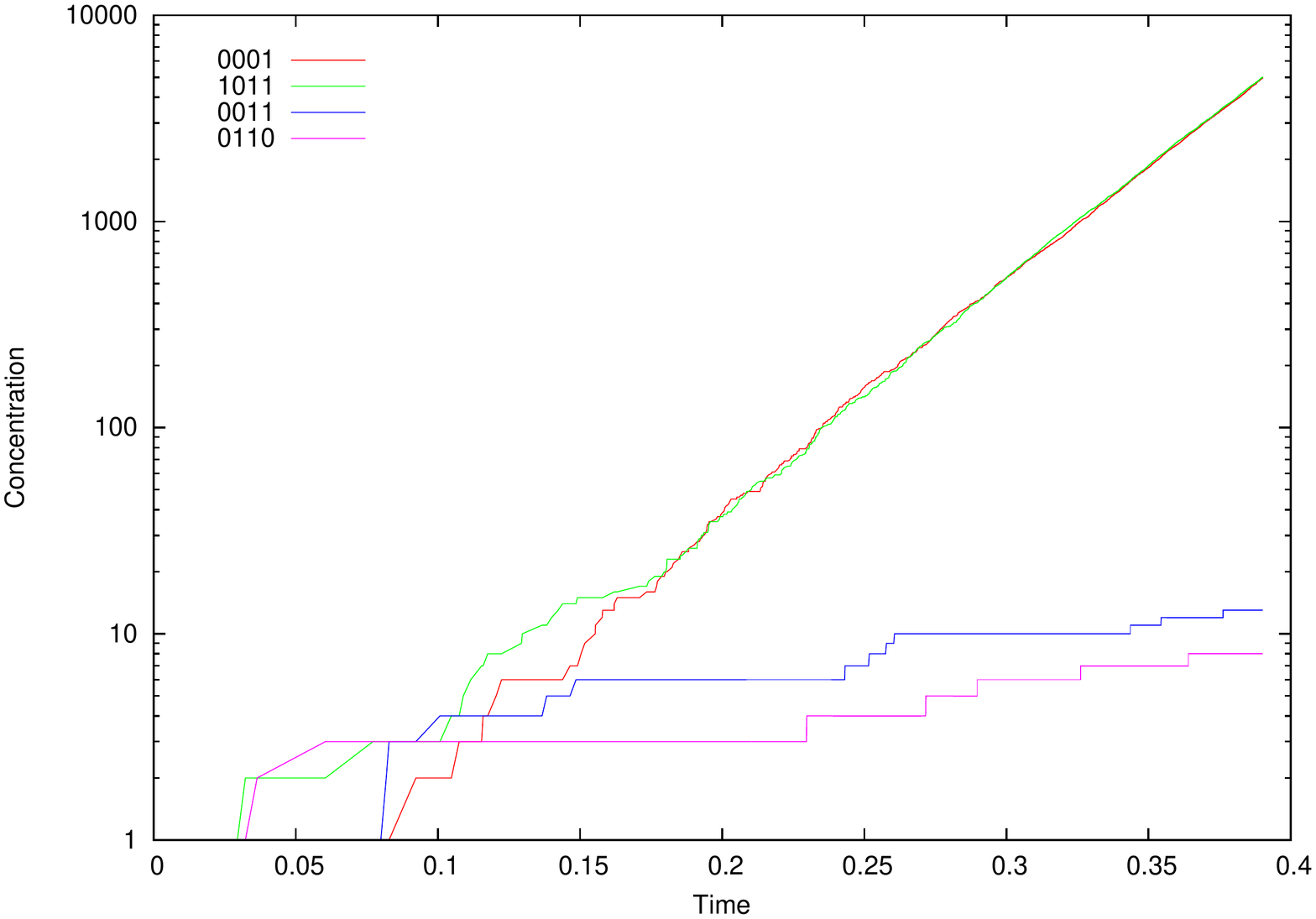}
\newpage
\subsection*{Figure 3 - Dynamics on the ligation and cleavage reaction graph}
The molecular concentrations over time for the products of the four reactions when both ligation and cleavage reactions are considered. The RAF set clearly has an advantage over the non-RAF set.\\
\includegraphics[width=15cm]{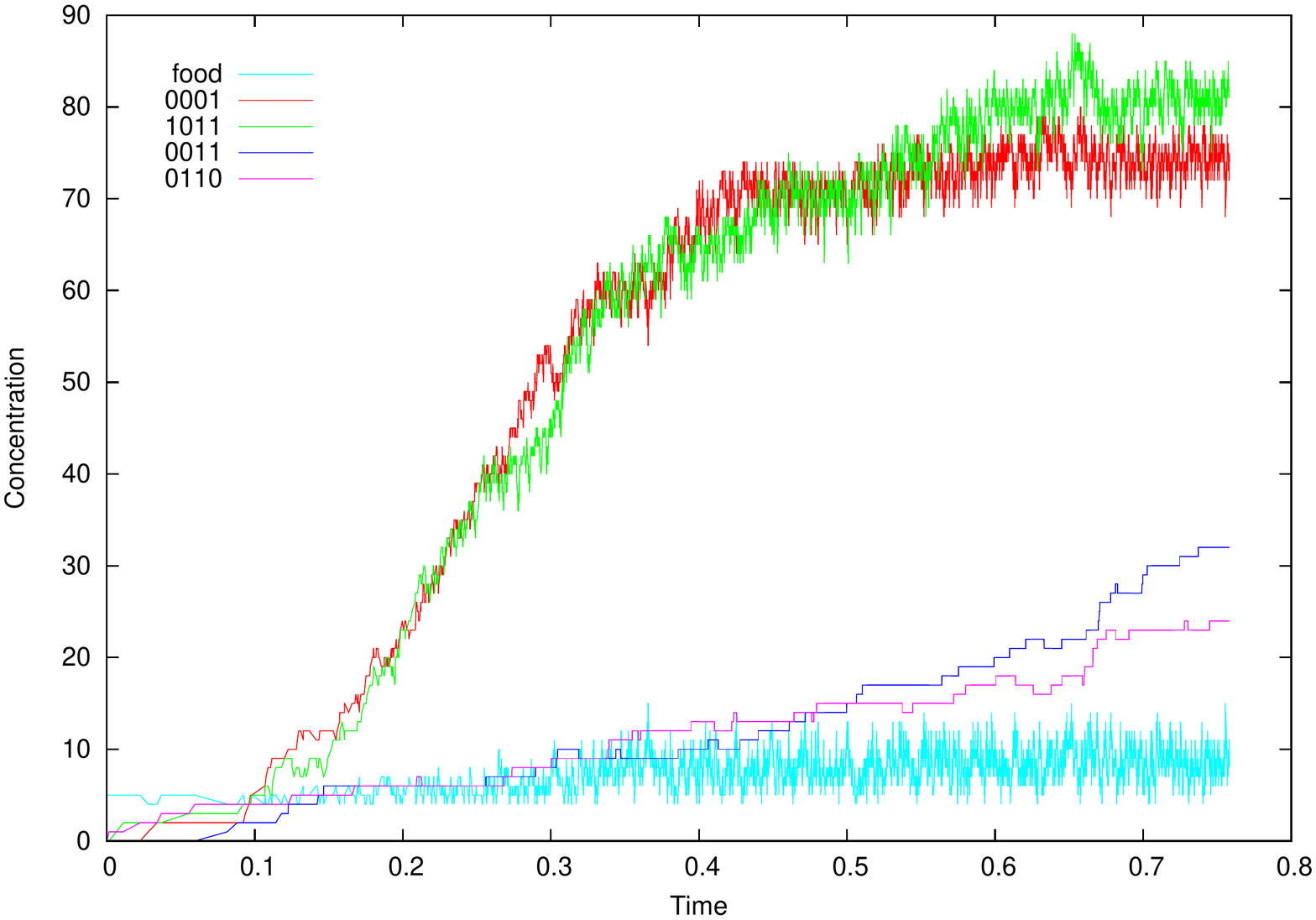}
\newpage
\subsection*{Figure 4 - Dynamics with low ligation to cleavage reaction ratio}
The molecular concentrations over time for the products of the four reactions with a low ligation to cleavage reaction rate ratio. The rate at which product molecules are broken down is too high for the RAF set to maintain an advantage over the non-RAF set.\\
\includegraphics[width=15cm]{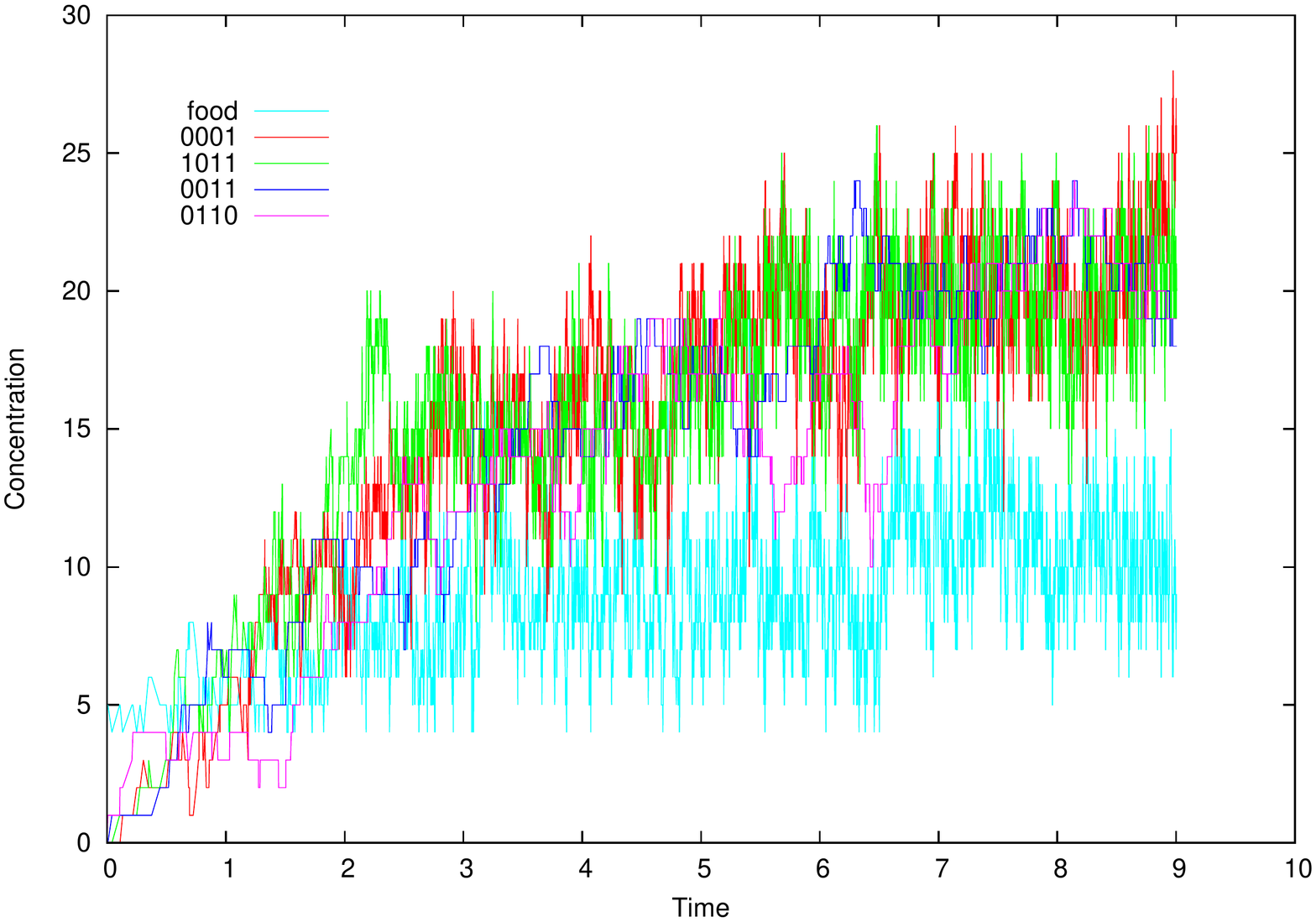}
\newpage
\subsection*{Figure 5 - A realistic RAF set}
A maximal RAF set as found by our RAF algorithm in an instance of the binary polymer model. The different subRAFs are indicated by colored boxes.\\
\includegraphics[width=10cm]{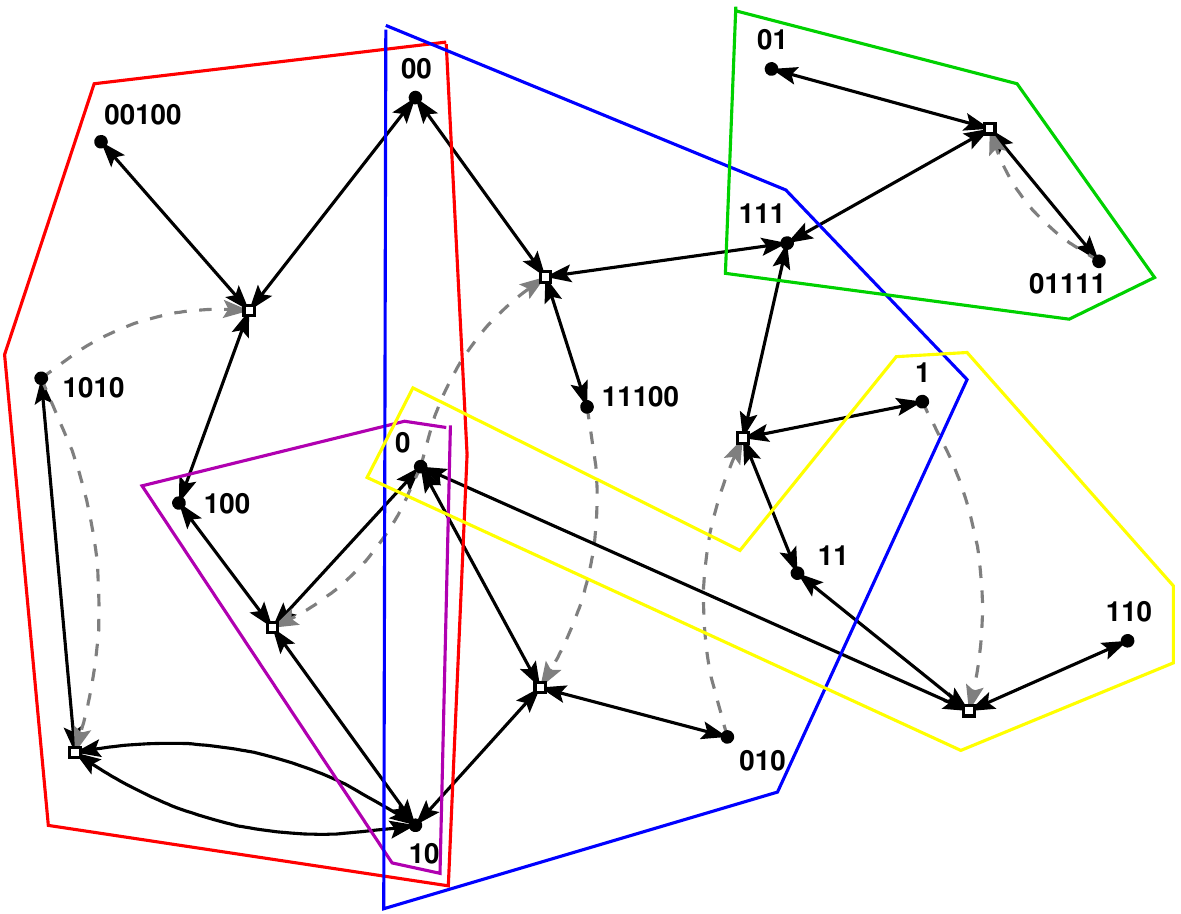}
\newpage
\subsection*{Figure 6 - Dynamics on the RAF set}
The molecular concentrations over time for the 8-reaction maximum RAF set.\\
\includegraphics[width=15cm]{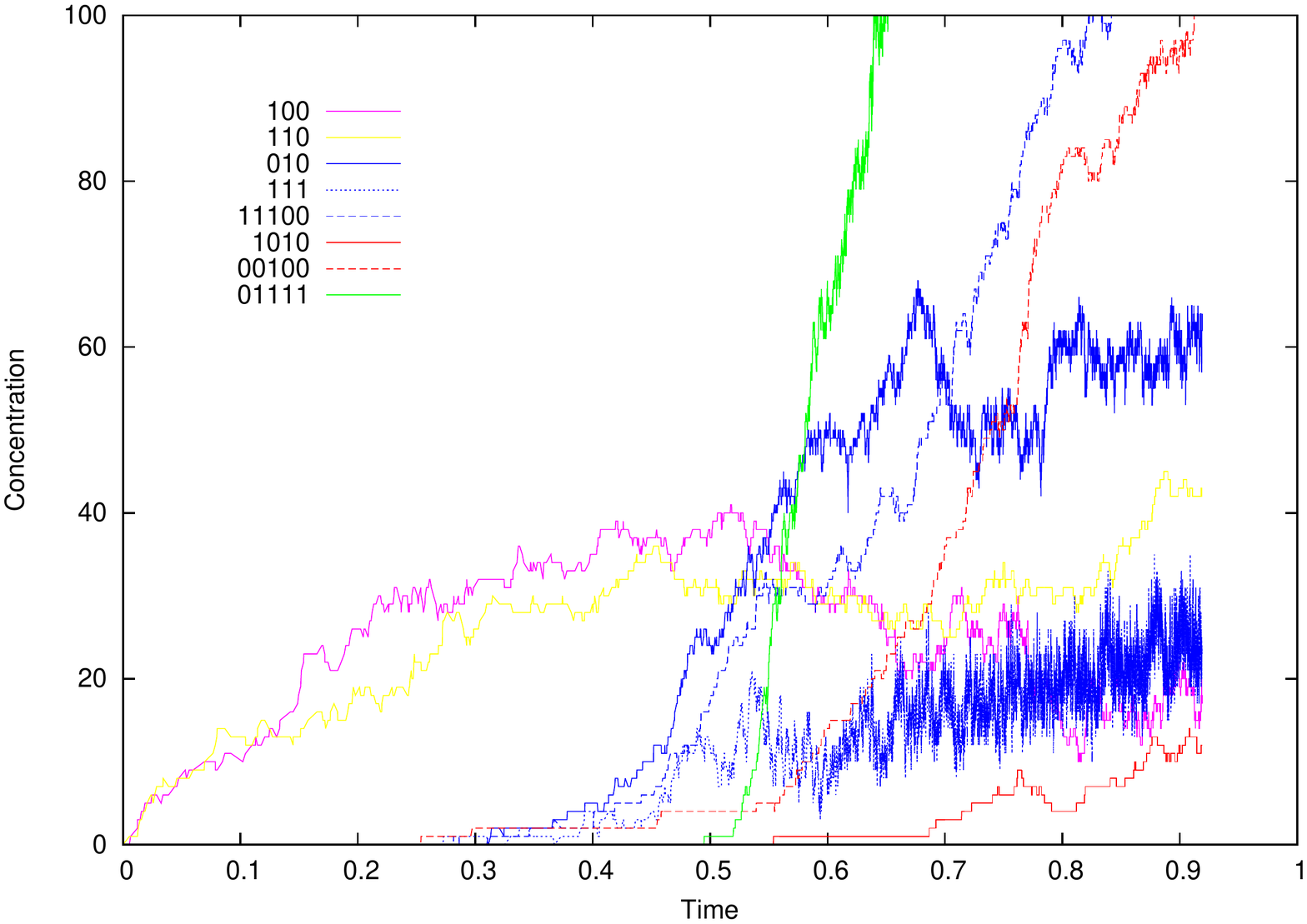}

\end{bmcformat}
\end{document}